\documentclass[a4]{epl2}
\usepackage{color}
\usepackage{amsmath}
\title{\textcolor[rgb]{0.00,0.00,0.00}{Field theoretic model for the Josephson effect}}
\shorttitle{\textcolor[rgb]{0.00,0.59,0.00}{Field theoretic model  for the Josephson effect}}
\author{A. I. Arbab\inst{}  \footnote{arbab.ibrahim@gmail.com}}

\institute{\inst{}
Department of Physics, College of Science, Qassim University, P.O. Box 6644,  Buraidah 51452,  KSA}


\abstract{The Josephson  effect is found to stem from the quantum behavior of massive photons existing in a superconducting medium. Accordingly, the Josephson coupling energy is found to be equal to the rest mass energy of these photons.  The Josephson effect is described by propagation of massive photon field following the universal quantum equation instead of being due to quantum tunnelling. The mass of the photon is found to be dependent on the electric properties of the junction. A  characteristic (critical) quantized capacitance of the junction is found to be  inversely related to the critical current.   The quantum (kinetic) inductance induced in the junction is found to be equal to $L_q=\mu_0\lambda_J$\,, where $\lambda_J$ is the Josephson penetration depth, and $\mu_0$ is the free space permeability.}
\begin{document}

\maketitle
\textwidth=16.5cm
\baselineskip=20pt
\section{\textcolor[rgb]{0.00,0.07,1.00}{Introduction}}
 Maxwell theory best describes the electromagnetic waves. The photon is the carrier of the electromagnetic interactions. It travels at speed of light in vacuum, but with less speed in a medium. In a conducting medium the electromagnetic wave is damped by an amount called a skin depth. Light also get refracted when enters a transparent medium as the photon speed becomes less than that in vacuum.  Therefore, photons in vacuum are described by a plane wave while in a conducting medium by a wavepacket that resembles a quantum particle. This makes the photon (mediator) between electrons massive, and consequently the electromagnetic interactions become short range. The wave of massive photon is longitudinal whereas that due to massless photon is transverse.

The mass nature of the photon is not a property of photon, but is a manifestation of the  environment in which the photon exists. This mass is characteristic of the medium in which the photon moves \textcolor[rgb]{0.00,0.07,1.00}{\cite{1}}. The more conducting medium, the big mass the photon will have. While the electromagnetic wave (massless photon) in vacuum has an intrinsic impedance, the de Broglie  matter wave (massive photon) will have an impedance that characterises  the medium in which it ravels. This impedance is found to be related to Hall residence \textcolor[rgb]{0.00,0.07,1.00}{\cite{klit}}.

 A massive photon is governed by the generalized telegraph equation (relatively undistorted wave) that is a damped Klein-Gordon equation \textcolor[rgb]{0.00,0.07,1.00}{\cite{1,uqe, tele}}. The telegraph equation governs the propagation of signals (voltage/current) in electric systems. The transport of this massive photon in a medium would mimic the motion of Cooper pairs in the Josephson junction \textcolor[rgb]{0.00,0.07,1.00}{\cite{cooper, jose}}. In Josephson junction Cooper pairs are assumed to tunnel across the junction barrier. The junctions capacitive and resistive properties are modelled in the Resistively and Capacitively Shunted Junction-model (RCSJ) that bestowed its DC characteristics \textcolor[rgb]{0.00,0.07,1.00}{\cite{pend, pend1}}. Josephson effect deals with matter aspect of the Cooper pairs and not the field one, where the pairs are thought to tunnel through the barrier. We present here a field theoretic model for this effect. This model is also equivalent to a telegraph equation that models an electric circuit consisting of an inductor, resistor and capacitor. This reveals that  the electromagnetic waves can be developed on the wire, and that wave patterns can appear along the line.

 A special telegraph equation (undistorted) is found to govern a spin - 0 quantum particle \textcolor[rgb]{0.00,0.07,1.00}{\cite{uqe}}. It describes the moving quantum  particle by a damped  oscillating wavefunction (wavepacket), \emph{i.e.}, $\psi=C\exp\,(-mc^2t/\hbar)\exp\,i\, (\pm \,kct-\vec{k}\cdot\vec{r})$, where $C$ is a constant, and $\vec{k}$ is the propagation constant. Moreover, the frequency of the wavepacket is complex and is given by $\omega=-\frac{mc^2}{\hbar}\,\,i\pm c\,k$. The group velocity of the wavepacket is equal to the speed of light, $\pm\,c$. This is unlike the ordinary case where the group velocity is greater than the speed of light. We should therefore treat Cooper pairs not as particles but as quantum wave (field) characterized by our damped Klein-Gordon equation \textcolor[rgb]{0.00,0.07,1.00}{\cite{uqe}}. However, these pairs are known to be described by a time-dependent Landau-Ginzburg equation \textcolor[rgb]{0.00,0.07,1.00}{\cite{ginz}}. Here the exponentially damped wave in one superconductor can extend to the second one, as far as the thickness of the barrier is made sufficiently thin \textcolor[rgb]{0.00,0.07,1.00}{\cite{super-tunn}}. In the standard quantum mechanics  the probability  decreases exponentially with distance in the classically forbidden region. We will show in this work that the Josephson effect can be described by a field theoretic aspect rather than a single particle aspect. It is thus governed by a field theoretic equation (e.g., Klein-Gordon equation), rather than a single particle equation ( e.g., Schrodinger equation).

 This paper is structured as follows: We introduce in Section 2 the concept of quantum current, inductance resulting from the motion of massive photons in a super(conductor). The quantum current is found to exist even when no voltage is applied. The magnetic potential energy is found to be quantized too. This system is shown to be equivalent to a quantum RL circuit, where $R$ is related to Hall's resistance \textcolor[rgb]{0.00,0.07,1.00}{\cite{klit}}. In Section 3 we show the analogy between our approach and the Josephson junction effect \textcolor[rgb]{0.00,0.07,1.00}{\cite{jose}}. We show in Section 4 that Josephson junction is equivalent to a resonant LCR circuit. Section 5 deals with a quantum mechanical description of the Josephson effect. We show in this section that the Josephson coupling energy is equal to the rest mass energy of the massive photon that mimics the Cooper pairs responsible for the superconducting state \textcolor[rgb]{0.00,0.07,1.00}{\cite{cooper,bardeen}}.  We end our paper with some concluding remarks.

\section{\textcolor[rgb]{0.00,0.07,1.00}{Quantum current, inductance and force}}
In a recent paper we have shown that the electric current associated with massive photon in a conductor is given by \textcolor[rgb]{0.00,0.07,1.00}{\cite{matter}}
\begin{equation}
I=\pm\frac{mc}{\mu_0\hbar}\,\phi_B\,,
\end{equation}
where $\phi_B$ is the magnetic flux, and $I$ is the induced electric current in a conductor. The current in Eq.(1) results from the motion of massive photon associated with moving electrons.
We have found recently that photons acquires a mass inside a conducting medium that is given by \textcolor[rgb]{0.00,0.07,1.00}{\cite{1}}
\begin{equation}
m=\frac{\mu_0\sigma\hbar}{2}\,,
\end{equation}
where $\sigma$ is the electric conductivity.
Therefore, the mass of the photon is a characteristic of the medium in which it travels. And since $\sigma$ is proportional to the density of electron of the medium,  the denser the medium the bigger the photon mass. It seems that electrons in the medium screen the photons from propagating inside the medium, and this compels the photon to acquire  mass.
The Compton (de Broglie) wavelength and frequency associated with the massive photon inside a conductor, which moves as a matter wave, are
\begin{equation}
\lambda _q=\frac{c}{K\,\sigma}\,,\qquad f_q=K\,\sigma\,,
\end{equation}
where $K$ is the Coulomb's constant.
Equations (1) and (2) permit us to define a self-inductance quantum
\begin{equation}
L_q=\pm\,\frac{\mu_0\hbar}{mc}=\frac{2}{\sigma\,c}\,,
\end{equation}
as a fundamental inductance associated with massive photons resulting from the motion of the magnetic charges.
Since the magnetic flux for superconducting phase is quantized in units of $\frac{\hbar}{2e}$  \textcolor[rgb]{0.00,0.07,1.00}{\cite{matter}} $\phi_B=n\frac{\hbar}{2e}$,  Eq.(1) yields
\begin{equation}
I_q=\pm\, \frac{mc}{\mu_0(2e)}\, n\,,\qquad n=0, 1, 2, \cdots.
\end{equation}
Hence, applying Eq.(2) in Eq.(5) gives
\begin{equation}
I_q=\pm\, \frac{c\sigma\hbar}{2(2e)}\, n\,.
\end{equation}
The current quantum, $I_q$, represents the quantum current that can be developed from magnetic charge moving inside super(conductors). It is interesting to see that $I_q$ is  quantized, while $L_q$ is not. Moreover,  the electric voltage is zero, \emph{i.e.,}
\begin{equation}
V=0\,.
\end{equation}
This is so because the electric field of massive photons vanishes inside a conducting medium.

 The magnetic energy deposited in the system is $U_q=\frac{1}{2}L_qI_q^2$ and upon applying Eqs.(4) and (6) becomes
\begin{equation}
U_q=\frac{c\,\sigma\,\hbar^2}{4(2\,e)^2}\, n^2\,\,.
\end{equation}
One can associate a quantum potential with Eq.(8) so that $V_q=\frac{c\,\sigma\,\hbar^2}{4\,(2e)^3}\, n^2=U_q/(2e)$.  Upon writing $V_q=I_qR_{q,n}$, and using Eq.(6), one finds that
\begin{equation}
R_{q,n}=\frac{\hbar}{2(2e)^2}\, n\,.
\end{equation}
This can be related to the Klitzing  resistance of the quantum Hall effect, $R_H=\frac{h}{e^2}$ \textcolor[rgb]{0.00,0.07,1.00}{\cite{klit}}.  It is still premature to give full account of the nature of the current quantum, inductance quantum, and force quantum. One can write from Eqs.(4), (6) and Eq.(8) the relations
\begin{equation}
U_q \left( \frac{8\alpha}{k\sigma}\right)=\frac{\hbar}{2}\,,
\end{equation}
and
\begin{equation}
L_q\,(eI_q)=n\frac{\hbar}{2}\,,
\end{equation}
where
\begin{equation}
 \Delta E=U_q\,,\,\,\qquad \Delta t=\frac{8\alpha}{K\,\sigma}\,,\,\qquad\, K=(4\pi\varepsilon_0)^{-1},\,\,\qquad \alpha=\frac{Ke^2}{\hbar c}\,.
\end{equation}
In the spirit of Eq.(11),  Eq.(10) represents the Heisenberg  uncertainty principle where $\Delta E$ represents the minimum possible energy that can be measured, and $\Delta t$ the time taken for this measurement. If we now equate $\Delta t$ to the inductive time of  RL circuit, $\tau_L=\frac{L_q}{R_q}$, one then finds $R_q=\frac{\hbar}{(2e)^2}$. This may represent the Hall resistance of the Cooper pairs \textcolor[rgb]{0.00,0.07,1.00}{\cite{hallsu}}. This is same as the value obtained from Eq.(9) with $n=2$. One can also associate the relation
$\tau=\hbar/(k_BT)$ with Josephson junction, where $T$ is the absolute temperature and $k_B$ is the Boltzman's constant. Moreover, since the junction is associated with tetrahertz frequency,  $T$ must be a few Kelvin.

\section{\textcolor[rgb]{0.00,0.07,1.00}{Standard Josephson junction effect}}

A Josephson junction can be defined as a superconductor-insulator-superconductor junction that  allows superconducting current  carried by  Cooper pairs through the junction. Cooper pairs can tunnel through the insulator without the presence
of a voltage, when the current through the junction is smaller than some critical value $I_c$. The current is due to the phase difference of the wavefunctions between the two sides of the junction. It is given by \textcolor[rgb]{0.00,0.07,1.00}{\cite{jose}}
\begin{equation}
I=I_c\sin\varphi\,.
\end{equation}
 This condition  mimics the situation in Eqs.(6) and (7), where $I_c=I_q$, with $n=1$.
 If the current through the  junction exceeds $I_c$,
an additional current of normal electrons flows. However, when a DC voltage is applied to the junction, the current oscillates with a frequency given by
\begin{equation}
\frac{\partial\varphi}{\partial t}=\frac{\hbar}{2e}\, V\,,
\end{equation}
so that the coupling energy
\begin{equation}
E=\frac{\hbar\,I_c}{2e}\,\cos\varphi\,.
\end{equation}
The Josephson  energy is connected with the junction critical current and inductance  by \textcolor[rgb]{0.00,0.07,1.00}{\cite{jose}}
\begin{equation}
I_c=\frac{2e}{\hbar}\,E_J \,,\qquad L_c=\frac{\hbar}{2eI_c}\,.
\end{equation}
 These Josephson equations suggest that $E_J (\frac{2e^2L_c}{\hbar})=\frac{\hbar}{2}$. This can be seen as defining an uncertainty relation where $\Delta E=E_J$ and $\Delta t=\frac{2e^2L_c}{\hbar}$. Hence, $E_J=2U_q$ when $n=1$. This may suggest that the Josephson energy is the total energy of moving Cooper pairs that have a kinetic energy equals to the potential energy. If we equate this time to the inductive  time of RL circuit, $\tau_L=\frac{L_c}{R_J}$, we will find that $R_J=\frac{\hbar}{2e^2}$ so that  $R_J=2\,R_q=R_H/(4\pi)$. Therefore, $R_J$ is  related to the standard Klitzing's resistance. Hence, one can treat Josephson and Hall resistances as equivalent to a quantum RL circuit. We have shown in a recent paper that the Hall's resistance can be thought of as the impendence that the electron experiences while moving in a conductor \textcolor[rgb]{0.00,0.07,1.00}{\cite{resist}}. Moreover, we see that $I_cL_c=I_qL_q$.

 The critical current for two identical superconductors at absolute 0K is given by \textcolor[rgb]{0.00,0.07,1.00}{\cite{ambog}}
 \begin{equation}
I_c=\frac{\pi \Delta (0)}{2e R_n}\,,
\end{equation}
where $R_n$ is the  normal state resistance of the tunnel junction, and $\Delta (0)$ is  superconducting energy gap at  $0K$. Comparing this with Eq.(6) reveals that the energy gap $2\Delta (0)=\hbar\,\omega_J$, where
$$\omega_J=\frac{c\,\ell}{\pi  A}\,,$$
where $\ell$ and $A$ are the length and the cross-sectional area of the junction, respectively. The frequency $\omega_J$ could be seen as some critical  frequency of the junction.

\subsection{Unruh radiation}
It is hypothesized by Unruh that an the effective temperature experienced by a uniformly accelerating detector in a vacuum field is given by \textcolor[rgb]{0.00,0.07,1.00}{\cite{unruh}}
$$T_U=\frac{a\,\hbar}{2\pi ck_B}\,,$$
where $a$ is the observation acceleration and $k_B$ is the Boltzman constant. Hence, if we assume the temperature of the junction to be due to such an effect, then the Cooper pair will be acceleration relative to some background reference.  If the power emitted by an accelerating Cooper-pair ($q$) follows that of the Larmor radiation \textcolor[rgb]{0.00,0.07,1.00}{\cite{larmor}},
$$P=\frac{1}{6\pi\varepsilon_0}\frac{q^2a^2}{c^3}\,,$$
then one finds
$$P_U=I_U^2R\,,\qquad\qquad R=\frac{2\pi}{3}Z_0\,,\qquad\qquad I_U=\frac{qk_B}{\hbar}\,T_U\,,\qquad\qquad Z_0=\mu_0c\,.$$
If the radiation by the Josephson junction is due to Unruh radiation, then $P_U$ is the power emitted by the Josephson junction at the temperature, $T_U$. We call the current, $I_U$, the thermal current, since is related to the temperature. The current and acceleration are related by the relation
$$a=\frac{2\pi c}{q}\, I_U\,.$$
The acceleration of  the Cooper pair through the junction due to the applied voltage $(V)$ is given by
$$a=\frac{qV}{m w}\,,$$
where $w$ is the thickness of the junction. Applying this in the above equation yields
$$I_U=\frac{q^2}{2\pi m c w}\, V\,.$$
This voltage-current relation suggests a resistance of the form
$$R_U=\frac{2\pi mcw}{q^2}\,.$$
If we, furthermore, assume that $\hbar\,\omega_J=k_BT_U$, then the Cooper-pair acceleration through the junction will be
$$a=\frac{2c^2\ell}{A}\,.$$
Applying this in the above equation yields
$$I_U=\frac{qc\,\ell}{\pi\,A}\,,\qquad\qquad a=\frac{2\pi c}{q}\, I_U\,,\qquad\qquad T_U=\frac{\hbar c\, \ell}{\pi k_BA}\,,$$
where $I_U$ could be related to the critical current of the junction.

The Unruh temperature will be given by
$$T_U=\frac{\hbar}{h_U}\,\frac{qV}{k_B}\,,\qquad\qquad h_U=2\pi mcw\,.$$
Here $h_U$ can be seen as the orbital angular momentum of the massive photon (Cooper pair). It is thus apparent that the smaller the width of the junction, the  higher the  radiation temperature.  Applying the above acceleration in the Larmor formula yields
$$P_0=\frac{V^2}{R_0}\,,\qquad\qquad R_0=6\pi \,\frac{Z_q^2}{Z_0}\,,\qquad\qquad Z_q=\frac{mc w}{q^2}\,,$$
where $R_0$ can be seen as a \emph{propagation resistance} inside the junction. Since this resistance is so big, the Unruh radiation is difficult to be a dominant contribution. Can we assign the Unruh radiation to gravitational radiation carried by massive bosons?

It is worth mentioning that a free quantum particle is found to exhibit an oscillation (Zitterbewegung) with a frequency $f_0=\frac{2mc^2}{\hbar}$ \textcolor[rgb]{0.00,0.07,1.00}{\cite{schrod}}. This could be attributed to a quantum force $F_q=\frac{2m^2c^3}{\hbar}$, giving a power  $P_q=\frac{2m^2c^4}{\hbar}$. When we equate this power to the maximal gravitational power, $P_m=(c^5/G)$, where $G$ is the gravitational constant, we obtain the definition for the Planck's mass, $m_p=(\frac{c\hbar}{G})^{1/2}$ \textcolor[rgb]{0.00,0.00,1.00}{\cite{maxmal}}.

\section{\textcolor[rgb]{0.00,0.07,1.00}{Quantum LCR circuit}}

Let us assume now that Josephson junction to be equivalent to a resonant LCR circuit, where the resistance has a damping effect, and  L and C resonate. The electrical potential energy in the capacitor is equal to the magnetic potential energy in the inductor ($\frac{1}{2}LI^2=\frac{1}{2}qV$, where $q=2e$), \emph{i.e.},
\begin{equation}
I=\frac{2e R_{Eq.}}{L_{Eq.}}\,.
\end{equation}
If we now set $R_{Eq.}=\frac{\hbar}{(2e)^2}$, Eq.(18) yields
\begin{equation}
L_{Eq.}=\frac{\hbar}{2e\, I_{Eq.}}\,\,,
\end{equation}
and the total energy of the system is given by
\begin{equation}
I_{Eq.}=\frac{2e}{\hbar}\,E_{Eq.}\,\,.
\end{equation}
It is interesting that Eqs.(19) and (20) are the same as Eq.(16) for Josephson junction. In this circuit the current and voltage differ by a phase angle $\alpha$, so that the total energy (power $\times$ time) is given by
\begin{equation}
U=I_{rms}V_{rms}\,\tau\,\cos\alpha\,,\qquad U=I^2_{rms}R\,\tau\,\cos\alpha\,,
\end{equation}
where $rms$ is the root-mean-square value, $\tau$ is the average time, $R=R_q$ and $I_q=I_{rms}$. If we now apply Eqs.(6) and (12) (where $\tau=\Delta t$) in Eq.(21), we will obtain
\begin{equation}
U=\frac{\hbar\,I_q}{2e}\, \cos\alpha\,\,,
\end{equation}
while the current is
\begin{equation}
I=I_{q}\sin(2\pi f_qt+\alpha)\,\,.
\end{equation}
Equation (22) and (23) are but the AC Josephson  equations, where $\alpha=\varphi$, and $U$ is the Josephson coupling energy \textcolor[rgb]{0.00,0.07,1.00}{\cite{jose}}. Hence, the phase difference between the two superconductors in Josephson junction is like the phase difference between voltage and current in RCL circuit. While the phase angle in AC-Josephson effect is time dependent, it is time-independent in RCL.

Using Eqs.(7) and (11), one can write the voltage ($2e\,V_q=hf_q$) as
\begin{equation}
V_q =\frac{n^2}{4\pi}\frac{h }{2e} f_q\,,\qquad f_q=\frac{\hbar\sigma\,c}{8e^2}\,.
\end{equation}
Moreover, using Eqs.(6) and (24) one finds
\begin{equation}
I_q =2e\, f_q\,n\,\,.
\end{equation}
The Josephson junction is found to show an AC effect with Shapiro steps at voltages given by $V_n=\frac{nh}{2e}\,f$, where $f$ is the microwave frequency \textcolor[rgb]{0.00,0.07,1.00}{\cite{shap}}. This may imply that $f_q=\frac{4\pi}{n}\,f$.
Equation (25) is very interesting as it shows that a particle carrier of twice electron charge (Cooper pairs) is involved in the current transmission.

If we use the relation $C_q=2e/\,V_q$, together with Eqs.(24) and (25), we will obtain the relations,
\begin{equation}
I_q  =\frac{2(2\,e)^3}{n\hbar}\frac{1}{C_q}\,,\qquad C_q=\frac{4(2\,e)^4}{c\,\hbar^2n^2}\,\frac{1}{\sigma}\,.
\end{equation}
Equation (26) states that the current $I_q$ depends only on the capacitance and nothing else. Since $I_q$ is quantized, then $C_q$ is quantized too. Equation (26) shows that if $I_q$ is measured, one can calculate $C_q$ and $\sigma$. It is interesting to see from Eq.(26) that the larger the conductivity the smaller the capacitance. The capacitance $C_q$ may express the junction capacitance.

Using Eqs.(26) and (3), one finds
\begin{equation}
\lambda_q =\frac{\pi}{\mu_0}\,\left(\frac{\hbar}{(2e)^2}\right)^2C_q\,n\,,\qquad \lambda_q=\frac{1}{\mu_0}\,\frac{h}{2e}\, \frac{n}{I_q}\,.
\end{equation}
Equation (27) clearly states that the Compton wavelength of the photon can be calculated if the current passing in the junction is measured.
Now if the junction is irradiated by electromagnetic wave with energy $E_q$, then the absorption of this wave by the junction will occur when $E_q=h\,f_q$, then Eqs.(6), (8) and (25) lead to energy - current relations
\begin{equation}
E_q =\frac{h}{2e}\,\frac{I_q}{n}\,,\qquad U_q=\frac{\hbar\,n}{4e}\,I_q\,.
\end{equation}
Moreover,
\begin{equation}
\frac{E_q}{U_q}=\frac{4\pi}{n^2}\,.
\end{equation}
At resonance, $\tau_C=\tau_L$ so that $R_c=\sqrt{\frac{L_q}{C_q}}=R_q\sqrt{\frac{2}{n}}$. This resistance equals to $R_q$ for $n=2$.

Let us write an expression for the capacitance per surface area ($C_s$) for a capacitor of width $d$, using Eq.(26), and the fact that $\frac{1}{\sigma}=\frac{R_qA}{d}$, where
\begin{equation}
C_s=\frac{C_q}{A}=\frac{2(2e)^2}{n\,c\,\hbar}\,\frac{1}{d}\,.
\end{equation}
For $d$ in Angstrom, one has
\begin{equation}
C_s=\frac{6.47}{n\, d}\,\rm \mu\,F/cm^2\,.
\end{equation}
\section{\textcolor[rgb]{0.00,0.07,1.00}{Quantum mechanical considerations}}
We assume here that the current in Josephson junction is due to a passage of a quantum field (particle) with mass $m$. And since the phase difference is related to the vector potential $\vec{A}$ by the relation $\varphi\equiv\Delta\theta=\frac{2e}{\hbar}\,\int\vec{A}\cdot d\vec{\ell}$, then the phase difference will also satisfy the same quantum equation as that of $\vec{A}$. The vector and scalar fields, $\vec{A}$ and $\phi$, of the electromagnetic field are shown to have  physical reality \textcolor[rgb]{0.00,0.07,1.00}{\cite{ahar}}. In our approach here, $\vec{A}$ and $\phi$ are the wavefunctions of the massive photons \textcolor[rgb]{0.00,0.07,1.00}{\cite{matter}}.
The vector potential $\vec{A}$ due to massive photon is found to satisfy the unified quantum equation \textcolor[rgb]{0.00,0.07,1.00}{\cite{uqe}},
\begin{equation}
\frac{1}{c^2}\frac{\partial^2\vec{A}}{\partial t^2}-\nabla^2\vec{A}+\frac{2m}{\hbar}\frac{\partial\vec{A}}{\partial t}+\frac{m^2c^2}{\hbar^2}\,\vec{A}=0\,.
\end{equation}
Equation (32) represents a quantum equation of a damped field $\vec{A}$. The scalar potential $\phi$ is governed by the same equation too. It describes a traveling wavepacket without distortion. It can be seen as a linear perturbed sine-Gordon equation  \textcolor[rgb]{0.00,0.07,1.00}{\cite{sine}}.  The magnetic field in the junction is related to $\vec{A}$  by $\vec{B}=\vec{\nabla}\times\vec{A}$ and satisfies Eq.(32) too. The general solution of Eq.(32) describes a field traveling to the left and write with same frequency and wavelength, and  is given by, $\vec{A}=\vec{A}_0\exp(-mc^2t/\hbar\, )\,\exp i(\pm\, ckt-\vec{k}\cdot\vec{r})$, where $\vec{A}_0$ is constant vector and $\vec{k}$ is the propagation vector. This is a wavepacket whose group velocity is equal to the speed of light ($\pm\,c$), and its phase velocity is equal to $v_p=E/p$, where $E$ is the total relativistic energy and $p$ is the momentum ($p=\hbar\,k$). Equation (32) is of a general telegraph equation governing the propagation of electric signals in telephone and telegraph lines.
Assuming the superconducting phase difference is spatially uniform and satisfies an equation similar to  Eq.(32), we obtain
\begin{equation}
\frac{1}{c^2}\frac{d^2\varphi}{d t^2}+\frac{2m}{\hbar}\frac{d\varphi}{d t}+\frac{m^2c^2}{\hbar^2}\,\varphi=0\,.
\end{equation}
Josephson junction when the phase is not uniform (space dependent) is  considered by Sven-Olof Katterwe, where the resulting equation is called the perturbed sine-Gordon equation (or damped Klein Gordon equation) \textcolor[rgb]{0.00,0.07,1.00}{\cite{phd}}. The linear equation corresponds to Eq.(32).
Equation (33) has a solution of the form $\varphi=C\exp\,(-mc^2t/\hbar)\,\cos \,(\,\omega t)$, where $C$ is a constant, and $\omega=ck$ is the angular frequency of oscillation.
Equation (33) can be compared with Josephson junction equation \textcolor[rgb]{0.00,0.07,1.00}{\cite{jose}}
\begin{equation}
\frac{\hbar C}{2e}\frac{d^2\varphi}{d t^2}+\frac{\hbar}{2e R}\frac{d\varphi}{d t}+I_c\sin\varphi=0\,,
\end{equation}
where $I_c$ is the critical supercurrent, $C$ is the junction capacitance, and $R$ is the resistance of the junction, when no external current is applied.
The damping occurs when a field passes through the insulator in the Josephson junction. It is shown that Eq.(34) is also analogous to the oscillation of a rigid simple pendulum \textcolor[rgb]{0.00,0.07,1.00}{\cite{pend, pend2}}. Let us now consider the linear solution ($\sin \varphi\approx \varphi$) and compare  Eq.(33) and Eq.(34) to obtain
\begin{equation}
RC=\frac{\hbar}{2mc^2 }\,,\qquad C=\frac{2e\hbar}{m^2c^4}\,I_c\,.
\end{equation}
Equation (35) can be rearranged to give
\begin{equation}
m=\frac{4eI_cR}{c^2}\,,\qquad \hbar=8 \,eI_cR^2C\,.
\end{equation}
Equation (36) reveals that the transport of inertial (de Broglie) wave, characterised by $m$ and $\hbar$, is equivalent to transport of electron (signal) in electronic system characterise by  $R, I$ and $C$. Furthermore, we can associate a kinetic inductance that results from the situation when $\tau_L=\tau_C$, \emph{i.e.}, $L_k=R^2C$. Using Eq.(36) this gives, $L_k=\hbar/(8eI_c)$ that can be compared with Eq.(16). The two systems express transport of electromagnetic energy carried out by the field (particle), but exhibited in different form. In this sense an inertial wave is inherently equivalent to RC circuit. The factor $D=\hbar/(2m)=c^2RC$ reflects the diffusivity nature of the propagating inertial wave or signal in the medium. The diffusion length can be defined as $l_D=\sqrt{D\tau}=\frac{\hbar}{2mc}$.

If we now let $R=R_q=\frac{\hbar}{2(2e)^2}n$, then Eq.(36) yields
\begin{equation}
m=\frac{\hbar}{2e}\,\frac{I_c}{c^2}\,n\,,
\end{equation}
and
\begin{equation}
C_c=\frac{(2e)^3}{I_c\hbar\,n^2}\,.
\end{equation}
Equation (37) shows that the mass is quantized, while Eq.(38) expresses a quantized critical capacitance of the junction. Hence, the mass of photon  and the critical capacitance of the junction can be calculated if $I_c$ is known. Equation  (37) is very interesting  as it relates the inertial properties of a particle to its electric properties. This relation could be applicable to photons only, and not to other particles. The mass in Eq.(37)  may be termed the electrical mass.  Equation (38) can be compared with Eq.(26) obtained with different conditions. Equation (37)  can be written as
\begin{equation}
I_c=\frac{2e}{\hbar}\,\frac{mc^2}{n}\,.
\end{equation}
Equation (39) gives the electric current associated with massive photons due to its inertial (mass) effect \textcolor[rgb]{0.00,0.07,1.00}{\cite{inductance}}.
Comparing Eq.(39) with Eq.(16) reveals that the Josephson energy
\begin{equation}
E_J=mc^2\,,\qquad E_n=\frac{mc^2}{n}\,.
\end{equation}
Applying Eq.(2) in Eq.(40) yields
\begin{equation}
E_n=\frac{\hbar\,\sigma}{2\varepsilon_0}\,\frac{1}{n}\,,
\end{equation}
where $\sigma$ represents the effective conductively of the junction. Thus, $\sigma=E_J/(Kh)$ is a measure of conductivity of the junction.
The Josephson penetration depth can be defined by $\lambda_J=\frac{\hbar}{mc}$ ($=2\,l_D$) so that Eqs.(6) and (36) yields
$$\lambda_J=\frac{1}{\sigma R}\,.$$
This defines the area per length of the junction. Alternatively, one can use Eq.(2) and (4) to obtain
$$\lambda_J=\frac{2}{\mu_0c\,\sigma }=\frac{L_q}{\mu_0}\,.$$
The above equation states that the inductance $L_q$ is a measure of  the Josephson penetration depth \textcolor[rgb]{0.00,0.07,1.00}{\cite{inductance}}.
The above two equations imply that $R=\frac{\mu_0c}{2}$. This is actually equal to half the vacuum impedance of the electromagnetic wave. It is indeed related to Klitzing resistance by the relation $R=\alpha\, R_H$ \textcolor[rgb]{0.00,0.07,1.00}{\cite{klit}}. This resistance may reflect the impedance that electrons experience when crossing the junction. Alternatively, one can say that the massless photon has the vacuum (intrinsic) impedance, and the massive photon has  half that value.

Equation (40) shows that the Josephson coupling energy reflects the rest mass energy of the photon inside the junction accompanying moving electrons. The pairing energy of Cooper pairs is rather low ($\sim 1\, meV$).

It is of interest to mention that in a recent paper we have found a maximum current stemming from the internal motion when a magnetic field is applied to a conductor that read \textcolor[rgb]{0.00,0.07,1.00}{\cite{resist, frac}}
\begin{equation}
I_0=\frac{e\hbar\, n_s}{\pi\,m}\,,
\end{equation}
where the current $I_0$ is a current without application of voltage, and $n_s$ is the number of electrons per unit area.

The inductance (for $n=1$) in Eq.(11), using Eq.(5) and (42), will be
\begin{equation}
L_q=\frac{2\pi\,m}{(2e)^2\,n_s}=\frac{\pi \mu_0\sigma}{n_s}\frac{\hbar}{(2e)^2}\,.
\end{equation}
This inductance can be related to the kinetic inductance that manifests the effect of inertial mass of a moving  charge when encounters alternating electric fields \textcolor[rgb]{0.00,0.07,1.00}{\cite{inductance}}. Using Eq.(47), one can related the London and Josephson penetration depths ($\lambda_L)$ and $\lambda_J$, respectively  by the relation
\begin{equation}
\lambda_J=\frac{2\pi}{\ell}\,\lambda^2_L\,,
\end{equation}
where $\ell$ is the length (thickness) of the junction. Thus, apart from the $\pi$ factor in Eq.(48), this relation is similar to Pearl length that characterizes the distribution of the magnetic field around the vortex center \textcolor[rgb]{0.00,0.07,1.00}{\cite{pearl}}. Using Eq.(27), a critical magnetic field density  can be written as ($B_c=\frac{\mu_0I_q}{2\pi\lambda_q}$)
\begin{equation}
B_c=\frac{1}{2\pi}\frac{\hbar}{2e}\frac{1}{\lambda_J^2}\,.
\end{equation}
This equation indicates that $\lambda_J$ represents the radius of the circular area in which a single quantum flux is contained.
\section{\textcolor[rgb]{0.00,0.07,1.00}{Concluding remarks}}
We have studied in this paper the consequences of having massive photons inside super(conductors). We have shown that Josephson and Hall effects are associated with the passage of massive photon (boson) in a conducting medium. The conductivity of the material can be calculated availing the interrelated measuring values. While Josephson explains the critical current as due to quantum tunneling effect, we interpret it as a flow of massive boson field through the junction.  It is thus a field effect rather than a single particle effect. The impact of massive photon could be manifested in several effects, such as Josephson junction. Moreover, the phase difference in the Josephson effect is found to be governed by our unified quantum equation. There exists a critical capacitance associated with massive photon in the junction. It is found that the massive photon has an impedance equals half that of the vacuum impedance for massless photon. The quantum inductance is found to be a measure of Josephson penetration depth. It also expresses the kinetic inductance of the superconducting state.
\section{\textcolor[rgb]{0.00,0.07,1.00}{Acknowledgements}}
I would like to thank Dr. M. Y. A. Yousif, H. Dirar, and M. A. H. Khalafalla for their stimulating discussion and encouragement.

\end{document}